\documentclass{bmcart}

\usepackage{amsthm,amsmath}
\usepackage{soul}
\usepackage{graphicx}
\usepackage{caption}
\usepackage{url}
\usepackage{subfigure}

\usepackage[utf8]{inputenc} 

\startlocaldefs
\endlocaldefs

\begin{document}

\begin{frontmatter}

\begin{fmbox}
\dochead{Research}


\title{DDeMON: Ontology-based function prediction by  Deep Learning from Dynamic Multiplex Networks}


\author[
  addressref={aff1},                   
  email={}   
]{\inits{J.K.}\fnm{Jan} \snm{Kralj}}
\author[
  addressref={aff1},                   
  corref={aff1},                       
  email={}   
]{\inits{B.\v{S}.}\fnm{Bla\v{z}} \snm{\v{S}krlj}}
\author[
  addressref={aff2},                   
  email={blaz.skrlj@ijs.si}   
]{\inits{Z.\v{R}.}\fnm{\v{Z}iva} \snm{Ram\v{s}ak}}
\author[
  addressref={aff1},                   
  email={}   
]{\inits{N.\v{L}.}\fnm{Nada} \snm{Lavra\v{c}}}
\author[
  addressref={aff2},                   
  corref={aff2},                       
  email={nada.lavrac@ijs.si},   
]{\inits{N.\v{L}.}\fnm{Kristina} \snm{Gruden}}

\address[id=aff1]{
  \orgdiv{Department of Knowledge Technologies},             
  \orgname{Jo\v{z}ef Stefan Institute},          
  \city{Ljubljana},                              
  \cny{Slovenia}                                    
}
\address[id=aff2]{%
  \orgname{National Institute of Biology},
  \city{Ljubljana},
  \cny{Slovenia}
}



\end{fmbox}


\begin{abstractbox}

\begin{abstract} 
Biological systems can be studied at multiple levels of information, including gene, protein, RNA and different interaction networks levels. The goal of this work is to explore how the fusion of systems' level information with temporal dynamics of gene expression can be used in combination with non-linear approximation power of deep neural networks to predict novel gene functions in a non-model organism potato \emph{Solanum  tuberosum}. We propose DDeMON (Dynamic Deep learning from temporal Multiplex Ontology-annotated Networks), an approach for scalable, systems-level inference of function annotation using time-dependent multiscale biological information. The proposed method, which is capable of considering billions of potential links between the genes of interest, was applied on experimental gene expression data and the background knowledge network to reliably classify genes with unknown function into five different functional ontology categories, linked to the experimental data set. Predicted novel functions of genes were validated using extensive protein domain search approach. \\
\end{abstract}


\begin{keyword}
\kwd{Multilayer networks}
\kwd{Complex networks}
\kwd{Machine learning on graphs}
\end{keyword}


\end{abstractbox}
%

\end{frontmatter}


\section{Introduction}

Fusing data sources connecting multiple aspects of a biological system can yield better, more reliable models, relevant for e.g., biomarker and drug target discovery \cite{lanckriet2004statistical,ye2008heterogeneous,fotakis2019neofuse}. Merging information from different levels of biological interaction is a non-trivial problem, which has been
explored for the case of cancer development \cite{pfeffer2019data}, Alzheimer's disease progression \cite{mohammadi2017structured} and similar diseases \cite{ZITNIK201971}.

With high-throughput technologies maturing, numerous genomes of organisms, including plants, are being sequenced. However, when it comes to plants, we know what these protein-coding genes do for approximately 40\% of Arabidopsis (\emph{Arabidopsis thaliana}) and 1\% of rice (\emph{ryza sativa}) \cite{rhee2014, lamesch2012}. If we take into account only the Arabidopsis genes with experimentally confirmed functions, then only 13\% remains \cite{lamesch2012}. For this reason, functional annotation of genes using ontologies and transfer of ontological annotations between species using orthology is extremely important. In plant science, two of the most used ontologies are MapMan \cite{thimm2004} and the Gene Ontology \cite{gene2004}.

Multi-modal data analysis approaches are becoming prevalent in computational biology and bioinformatics, resulting in better and more robust models. Multi-modal biological data are often represented in network format, allowing for information from genomic, proteomic and other cellular layers to be simultaneously considered. Network-based approaches were successfully used e.g., for biological function prediction \cite{vascon2018protein,fan2016mirnet} and  characterization of structural domains \cite{vskrlj2018insights}. Complex networks are computationally and conceptually suitable for systems-level data integration, as demonstrated e.g., by the BioMine exploration tool \cite{podpevcan2019interactive}, representing the one of the largest integrated heterogeneous networks to date. 
While the existing approaches successfully address the problem of data fusion and the consequent learning for model organisms such as \emph{Homo Sapiens} and \emph{Mus Musculus}, development of approaches suitable for non-model organisms, such as crop plants, remains an open research problem.

This work presents an approach to modeling plant gene function using both static and \emph{temporal} information, incorporating multiple levels of biological information, including protein-protein, protein-gene and gene-gene interactions. 
Section~\ref{sec:related} discusses the related work. Section~\ref{sec:ddemon} presents the proposed DDeMON methodology. Section~\ref{sec:ddemon-prediction} addresses the problem of gene function prediction. We present the experimental results in Section~\ref{sec:results}, followed by the discussion and conclusions.

\section{Related work}
\label{sec:related}

\subsection{From complex to multiplex networks}
Networks can consist of multiple information layers. For example, the same entity (e.g., a protein coding gene) can be studied on the DNA-DNA, DNA-RNA, RNA-RNA and protein-protein interaction levels. Majority of currently known approaches only focus on a certain level (e.g., protein-protein interaction prediction).
Individual layers of interacting entities are commonly formalized as \emph{networks}. In its simplest form, a given network $G = (N,E,w)$ is comprised of a set of nodes $N$, a set of edges $E \subseteq N \times N$ and a weight function $w$ that assigns a real value to each edge.
Note that if the edges are directed, the edge sets are tuples (fixed order).
Such abstraction is, for example, suitable for modeling protein-protein interactions with varying degrees of empirical evidence (edge weights), or gene-gene correlations. 

\begin{figure}[b!]
    \centering
    \includegraphics[width=0.4\linewidth]{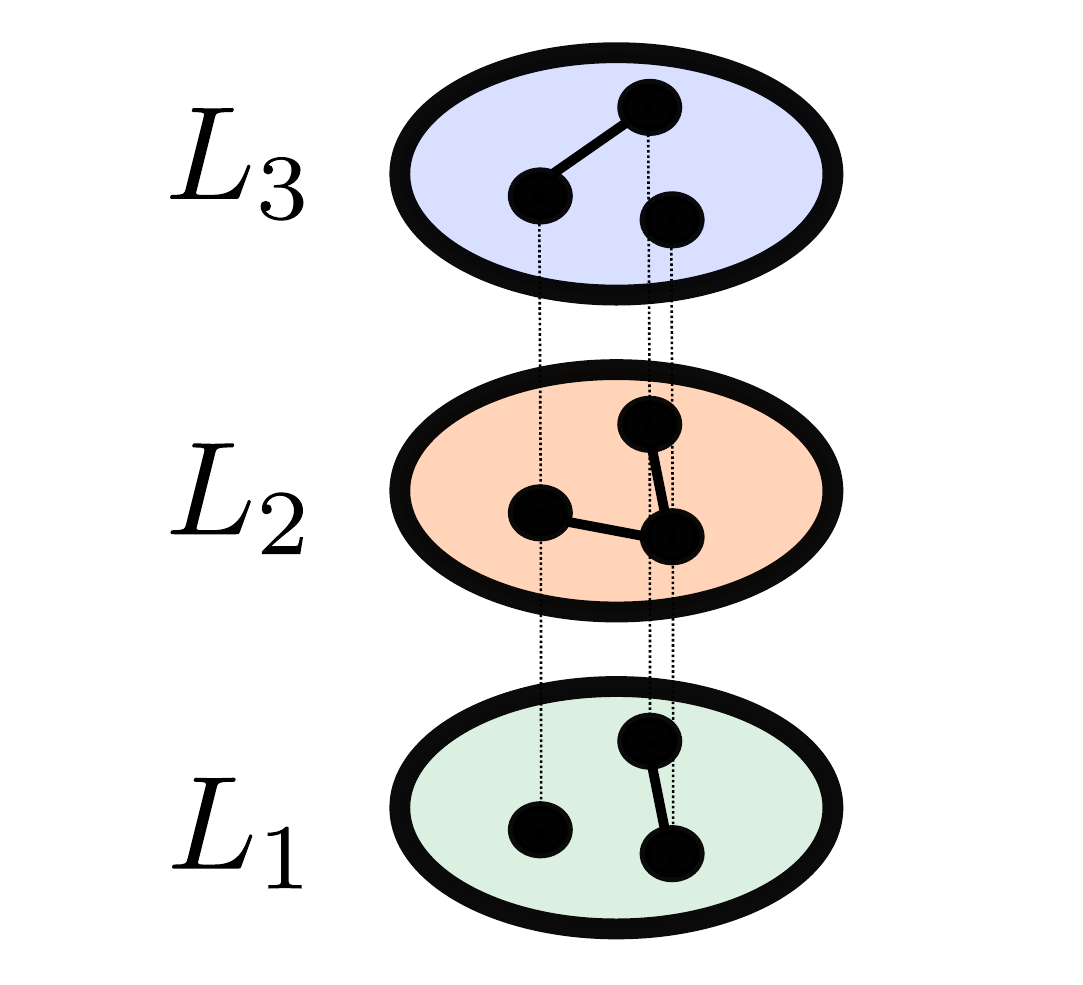}
    \caption{An example three-layered multiplex network with three nodes. The dotted vertical lines can be understood as the \emph{is\_a} relation. Note that even though the set of nodes is the same across all layers, the interactions between them differ.}
    \label{fig:multiplex-demo}
\end{figure}

When considering biological systems, however, the same set of e.g., genes can be studied at different levels of cellular function in terms of their co-expression, proteins they code for, homologs they map to and other \emph{modalities}. The notion of weighted networks defined above is not sufficient for representing such \emph{layered} structures. In this work we resort to the notion of \emph{multiplex networks} \cite{kivela2014multilayer} to capture the additional information levels, all within the same abstraction. Multiplex networks were successfully used to model biological systems, ranging from brain networks \cite{sola2013eigenvector} to metabolic networks \cite{didier2015identifying,valdeolivas2019random}.
A multiplex network $M$ can be defined as $M = \{G_i\}_i = \{(N_i,E_i,w_i)\}$, where $N_i = N_j$ for a given pair of layers $i$ and $j$. This structure associates the same set of nodes $N$ with different (weighted) edge sets $\{E\}_i$. For each layer, a separate set of edges $i$ is considered. 
A multiplex network spanning multiple layers, where each layer is a weighted, either directed or and undirected network, is the core abstraction underlying the proposed DDeMON methodology. A three layer multiplex network where three nodes interact differently across layers is shown in Figure~\ref{fig:multiplex-demo}.

\subsection{Dynamic multiplex networks}
\label{sec:dynamic-multiplex}

Considering different layers of information offers insights into different aspects of the same system, simultaneously. However, taking into account that e.g., expression profiles can be time-dependent, remains a lively research area.
For example, temporal multiplex networks were considered for modeling social phenomena \cite{starnini2017effects} as well as uncovering communities in time \cite{jiao2018constrained}.

When considering multiplex networks, either the whole network is dependent on time, or only parts of it (distinct layers). In this work we are primarily interested in the second situation, as one of the options for considering a given node's temporal behavior is to compare it with the remainder of the nodes' time series, yielding a new network representing the temporal relations between the nodes. Hence, the dynamics of single-layer homogeneous networks is being considered. We refer the interested reader to \cite{li2017fundamental} for a detailed overview of methods for analysis of temporal complex networks.

\subsection{Heterogeneous information networks}

\begin{figure}[t!]
    \centering
    \includegraphics[width=0.6\linewidth]{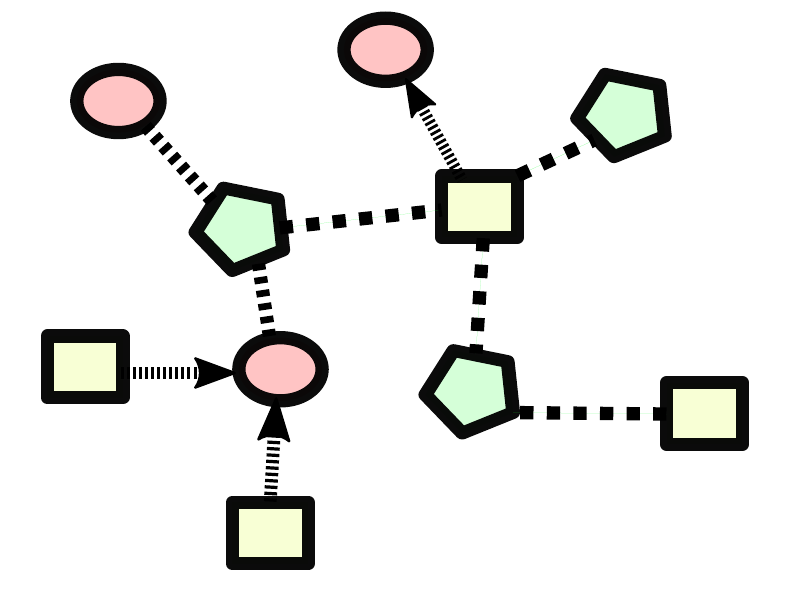}
    \caption{An example heterogeneous network. The network consists of three node types (circles, squares and pentagons). Each pair of distinct nodes (w.r.t. type) is connected with a different type of edge.}
    \label{fig:heterogeneous-demo}
\end{figure}

Some of the information layers, such as the associations between e.g., genes and the corresponding PubMed\footnote{https://pubmed.ncbi.nlm.nih.gov/} articles are not (weighted) networks on their own, and consist of multiple node and edge \emph{types}. Such networks are commonly referred to as \emph{heterogeneous information networks}.\footnote{Note that multiplex networks can be considered as a type of heterogeneous networks (different edge types between the same node type), however, for the purpose of this work, we maintain the two abstractions separate as the final network that is constructed by DDeMON is indeed a multiplex network.}  An example heterogeneous network is depicted in Figure~\ref{fig:heterogeneous-demo}.

The study of heterogeneous networks has yielded novel results across multiple disciplines \cite{schreiber2014heterogeneous}. An example use of this abstraction for representing diverse biological data bases is the BioMine explorer \cite{podpevcan2019interactive}, where different biological entities are linked across contexts. 
Heterogeneous networks need to be considered when constructing layers such as literature-based associations.

To this point, we discussed the key ideas considered for constructing a single multiplex network, where different aspects, ranging from literature-based associations to temporal relations between genes can be simultaneously taken into account. We next discuss the methodology capable of leveraging such rich structure in a machine learning setting.

\subsection{From heterogeneous to homogeneous networks}
\label{subs:HINMINE}

In the early 2000s, the notion of data fusion emerged  when researchers realized that the abundance of information available from different sources and contexts could---when merged e.g., into a single matrix---offer better means for modeling biological systems. 
Our early approach to data fusion for heterogeneous information networks \cite{grcar} uses differential evolution \cite{differentialevolution} to produce a weighted concatenation of the feature vectors constructed from different parts of a heterogeneous network.
Building on this work, we developed the HinMine methodology \cite{kralj2018} that offers a computationally more effective way of converting typed paths between the nodes of interest (e.g., genes) into weighted edges, using simple concatenation of feature vectors arising from different parts of a heterogeneous network. 
Following HinMine, the proposed DDeMON methodology employs the process of transforming heterogeneous networks to \emph{homogeneous} networks---i.e. weighted networks with a single node type (e.g., gene). Thus, if a given layer is not already a homogeneous network, it is transformed into one. Applied to all heterogeneous layers, the resulting structure is a standard multiplex network, representing e.g., different \emph{biological contexts}.

\subsection{Deep neural networks}
Computational neural network models were explored already at the end of the previous century \cite{rumelhart1995backpropagation,kennedy1988neural}. Recent advancements in hardware capabilities offered the opportunity to scale neural network models to larger data sets; for example images and large text corpora. The notion of deep learning refers to using neural network models, which consist of multiple hidden layers, offering better generalization capabilities as they have more potential to capture latent patterns, relevant to the problem at hand \cite{lecun2015deep}.
One of the most widely used types of neural networks are feedforward neural networks, which can be understood as stacked linear layers linked with non-linear activation functions. For example, if $\boldsymbol{X}$ represents the input data matrix and $\boldsymbol{W}_i$ denotes the weight matrix ($i$-th hidden layer), followed by a non-linear activation $a_i$ ($i$-th layer), one can represent a two layer neural network as:
$    l_o = a_2((a_1 (\boldsymbol{X}^T \cdot \boldsymbol{W}_1 +\boldsymbol{b}_1))^T \cdot \boldsymbol{W}_2 + \boldsymbol{b}_2),
$
\noindent where $l_o$ represents the output prediction(s) and $\boldsymbol{b}_i$ bias vectors.

Deep neural networks have been in the last decade successfully applied in plethora of biological domains. For example, graph-neural networks, a variant of neural networks capable of taking adjacency structure of (homogeneous) networks into account, were successfully applied for biomarker discovery and phenotype prediction \cite{li2019graph}. Further, neural network-based approaches were successfully adapted for the task of small molecule generation \cite{segler2018generating,simonovsky2018graphvae}, opening many new research directions for drug design and development.
Recently, first attempts at learning from highly heterogeneous, layered structures, such as the one considered by DDeMON were also explored \cite{wang2020abstract}. However, to our knowledge, learning from multiplex networks in a scalable manner remains an open research challenge  addressed in this work. 

\section{DDeMON: Constructing feature vectors describing target genes}
\label{sec:ddemon}

\begin{figure*}
    \centering
    \includegraphics[width=\linewidth]{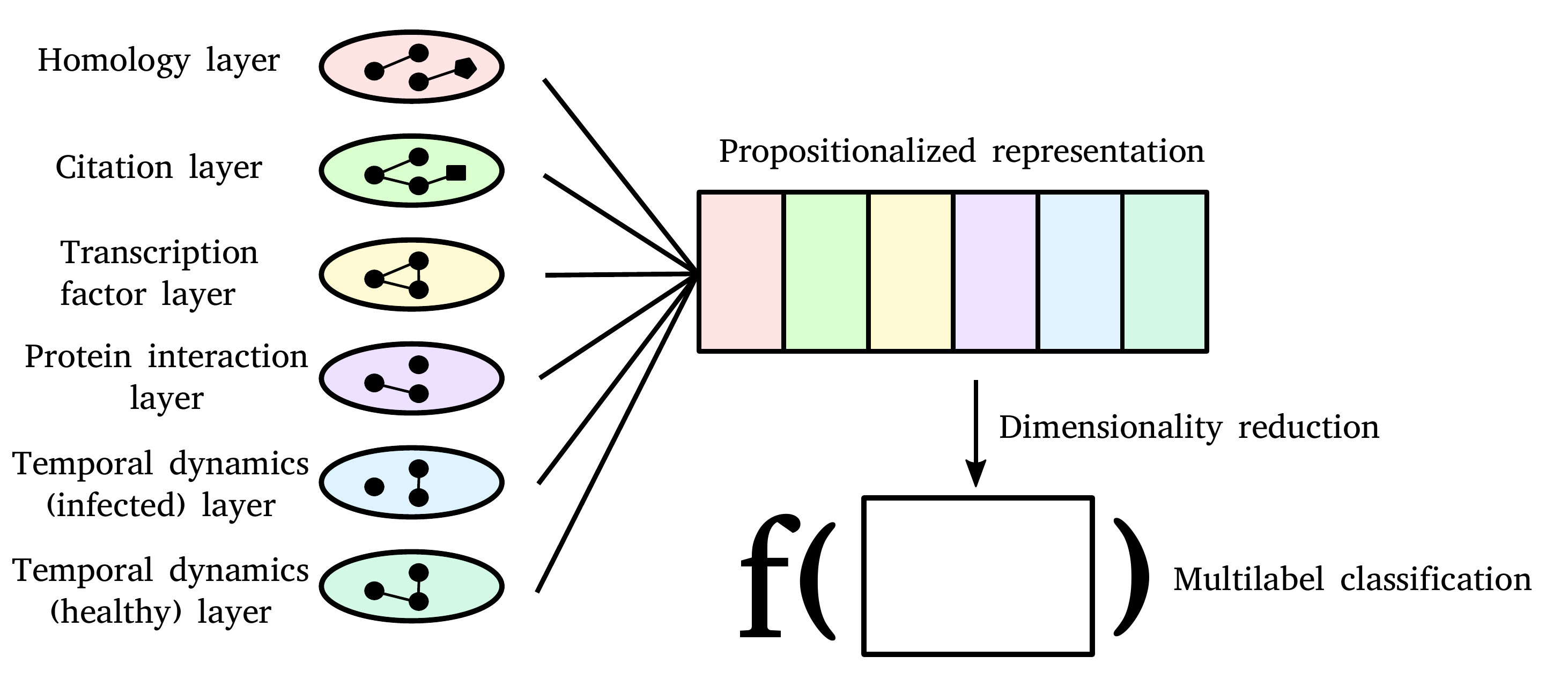}
    \caption{Overview of DDeMON. The approach consists of three main steps. First, a multiplex network is constructed. Here, if needed, individual heterogeneous networks (layers) are transformed to homogeneous ones (dotted lines denoting identity relation are omitted for readability). Next, the representation is generated only for the selected node type (circle), which are genes in DDeMON. The obtained representation is very high dimensional, and needs to be reduced to a lower-dimensional form suitable for learning, which is denoted with function $f$.}
    \label{fig:workflow}
\end{figure*}

This section presents the proposed DDeMON approach to learning from dynamic multiplex networks, which is outlined in Figure~\ref{fig:workflow}.

\subsection{Data used}
\label{subs:data}

We used data from several sources. First, a comprehensive knowledge network \cite{ramvsak2018}
was constructed by combining the graph of binary PIS-v2 interactions with three layers of publicly available information: protein-protein interactions (PPIs), transcriptional regulation (TR), and regulation through microRNA (miRNA). This resulted in an Arabidopsis thaliana comprehensive knowledge network with 20,012 nodes (19,812 genes, 186 miRNA families, three metabolites, and 11 viral proteins) and 70,091 connections. Each data layer covers unique gene or miRNA subsets in the entire network, with only six nodes present in all four layers, which indicates that our layer selection was well suited for inclusion. 

Second, we used data on gene expressions of 49,804 genes\footnote{\url{https://www.ncbi.nlm.nih.gov/geo/query/acc.cgi?acc=GSE58593}, \url{https://www.ncbi.nlm.nih.gov/geo/query/acc.cgi?acc=GSE46180}}. Briefly, samples were taken from fully developed potato leaves, that were treated either with sap of healthy plants (healthy samples) or sap of infected plants (infected samples). The transcriptional response was analysed in 4 different genotypes: Desiree, NahG-Desiree, Rywal and NahG-Rywal. Samples were taken at different times after treatment, creating a time series of 3 to 11 time points depending on the studied genotype.

Finally, we used the data from the GoMapMan website to assign genes to functional categories (MapMan ontology BINs) and publicly available PubMed data to connect the studied potato genes to publications mentioning or studying those genes.

\subsection{Multiplex network construction}
\label{sec:construction}

The first step of the DDeMON approach consists of heterogeneous network construction. We used the three data sources, described in Section \ref{subs:data}, to construct a single heterogeneous network. The basic statistics of the considered networks are given in Table~\ref{tab:my_label}. The types network nodes and edges are summarized in Tables~\ref{tab:nodes} and \ref{tab:edges}. 

\begin{table}[h!]
\vspace*{-0.5cm}
    \centering
        \caption{Basic statistics of the considered networks.}
    \label{tab:my_label}
    \begin{tabular}{lc} 
    \hline
    Network & Density\\
    \hline
        Gene-Paper-Gene & $0.00326$ \\
        CKN - Activation TF & $6.26 \cdot 10^{-6}$ \\
        CKN - Binding TF & $2.638\cdot 10^{-5}$\\
        CKN - Inhibition TF & $1.802\cdot 10^{-6}$\\
        CKN - unknown TF & $1.928\cdot 10^{-6}$ \\
        \hline
    \end{tabular}
    \vspace*{-0.5cm}
\end{table}

\begin{table}[h!]
\vspace*{-0.5cm}
\centering
\caption{Nodes of the heterogeneous network, constructed by DDeMON.}
\label{tab:nodes}
\begin{tabular}{lc}
\hline
Node type & Number of nodes\\\hline
Potato Gene (SoTub) & $49{,}804$\\
Arabidopsis Thaliana Gene (AT) & $33{,}341$\\
PubMed Article & $22{,}200$ \\
\hline
\end{tabular}
\vspace*{-0.5cm}
\end{table}

\begin{table}[h!]
\vspace*{-0.5cm}
\centering
\caption{Edges of the heterogeneous network, constructed by DDeMON.}
\label{tab:edges}
\begin{tabular}{lrc}
\hline
Edge type (node types) & Number of edges  & Directed \\
\hline
 Homolog-to (SoTub - AT) & $1{,}164{,}672$ & No\\
 Cited-by (AT - PubMed) & $135{,}737$ & Yes \\
 Transcription Factor-of (SoTub - SoTub) & $11{,}506$ & Yes\\
 Binding-to (SoTub - SoTub) & $30{,}382$ & Yes\\
 Experimental - healthy (SoTub - SoTub) & $1{,}151{,}719{,}969$ & Yes\\
 Experimental - infected (SoTub - SoTub) & $1{,}151{,}719{,}969$ & Yes \\
 \hline
\end{tabular}
\vspace*{-0.5cm}
\end{table}

Construction of separate network aspects/contexts is described below.

\noindent \textbf{Constructing \emph{Homolog} and \emph{Cited-by} edges.}
In order to construct the (undirected) edges of type \emph{Homolog} connecting potato genes with Arabidopsis  genes, we connected each potato gene to all Arabidopsis genes in the same homolog group. To obtain homolog groups for the genes, we used the data available on the GoMapMan website
\cite{10.1093/nar/gkt1056}.\footnote{http://protein.gomapman.org/export/current/biomine/ath\_homolog}
To then connect the Arabidopsis Thaliana genes to PubMed articles by directed edges, we used the TAIR database \cite{tair}\footnote{https://www.arabidopsis.org/} and converted the exports from TAIR into the network. 

\noindent \textbf{Constructing \emph{Transcription Factor} and \emph{Binding} edges.}
The edges of the type \emph{Transcription Factor} (either \emph{inhibition}, \emph{activation} or \emph{unknown}) and \emph{Binding} were extracted from a previously constructed comprehensive knowledge network (see above). 
For the purpose of our approach, we extracted \emph{all} the edges from the comprehensive knowledge network into our heterogeneous network. Note that the \emph{Transcription factor} edge type can be split further into three different edge types: \emph{inhibition\_TF}, \emph{activation\_TF}, \emph{other\_TF}.

\noindent \textbf{Constructing \emph{experimental} edges.}
The final set of edges we constructed were edges, induced directly from the gene expression data, described in Section \ref{subs:data}. For the purpose of DDeMON, we view the experimental data as a collection of time series, each time series charting the strength of the expression of one particular gene. 
To transform the data into a network, we used the time series data to induce weights on edges between each pair of genes. The goal was to construct a network where two genes are connected by a strong weight if the experimental data shows they share a similar expression profile (i.e. if the time series, describing their expression over time, are similar).

In order to determine the similarity between two time series, we used Dynamic Time Warping (DTW), the most widely used algorithm designed introduced in the 1960s \cite{wang1997alignment} for measuring similarity between two temporal sequences, which has since been used for various data analysis tasks on time series such as clustering or classification \cite{senin2008dynamic}. The algorithm  takes as input two time series $s=(s_1,s_2,\dots, s_n)$ and $t=(t_1, t_2,\dots t_m)$ of possibly different lengths, and a distance function $d$ measuring the distance $d(s_i, t_j)$ between elements of the series (typically, and in our example, the absolute difference between two numbers is used for $d$). The algorithm then finds a ``matching'' between elements of $s$ and elements of $t$. The matching $M\subseteq\{(s_i, t_j)|1\leq i \leq n, 1\leq j \leq m\}$ is a set of matches, i.e. ordered pairs $(s_i, t_j)$, and is uniquely defined by the following characteristics:
\begin{enumerate}
    \item $\forall i\exists j: (s_i, t_j)\in M$, meaning every element of $s$ is matched to at least one element of $t$.
    \item $\forall j\exists i: (s_i, t_j)\in M$, meaning every element of $t$ is matched to at least one element of $s$.
    \item $(s_0, t_0)\in M$, meaning the first element of $s$ is matched to the first element of $t$ (but it can also be matched to more elements of $t$).
    \item $(s_n, t_m)\in M$, meaning the last elements of the series are matched (but they can be matched to more elements as well)
    \item $(s_i, t_j)\in M \land i' > i, j'<j\implies (s_{i'}, t_{j'})\notin M$, meaning that the matching is ``increasing'' in $s$ - if $s_i$ is matched with $t_j$, then no element \emph{after} $s_i$ may be matched with an element \emph{before} $t_j$.
    \item $(s_i, t_j)\in M \land j' > j, i'<i\implies (s_{i'}, t_{j'})\notin M$, meaning the matching is increasing in $t$, as above.
    \item The sum 
    $\sum_{(s_i, t_j)\in M}d(s_i, t_j)$
    is minimized.
\end{enumerate} 

In DDeMON, for each pair of genes $g_1$ and $g_2$, we use the inverse value of the distance between their respective expression data series as the weight of the edge between the genes. The inverse value ensures that genes with more similar expression profiles will be connected by a stronger weight. As the raw data contains both expression profiles for genes in infected plants and in healthy plants, we repeat the same procedure on both parts of the raw data, thus constructing both \emph{Experimental\_healthy} and \emph{Experimental\_infected} edges in the network.

\subsection{Homogeneous network construction}

The second step of the DDeMON methodology is the construction of homogeneous networks comprised only of nodes, representing potato genes. This step is performed by decomposing the original heterogeneous network using a method also used by the previously developed HinMine algorithm (see Section~\ref{subs:HINMINE}). The step results in a set of homogeneous networks that all share the same set of nodes $V$, but have different sets of (possibly weighted) edges.

In each homogeneous network, two nodes are connected if they share a particular direct or indirect link in the original heterogeneous network. In particular, to construct the `\emph{connected via PubMed}' homogeneous network, we used the HinMine decomposition methodology to construct a network of genes in which two potato genes ($g_1$ and $g_2$) are connected if they are homologs to two Arabidopsis Thaliana genes ($at_1$ and $at_2$), mentioned in the same PubMed publication $p$. In other words, potato genes $g_1$, $g_2$ are connected if there exists a path from $g_1$ - \emph{Homolog-to} - $at_1$ - \emph{Cited-by} - $p$ - \emph{Cites} - $at_2$ -\emph{Homolog-to} - $g_2$. Note that the construction of homogeneous networks uses both the \emph{Cited-by} relation, described in Section~\ref{subs:data}, and the \emph{Cites} relation. The latter is defined as the reverse of the former, i.e. for a gene $g$ and paper $p$, we define that $p$ \emph{Cites} $g$ if and only if $g$ is \emph{Cited-by} $p$.

Analogously to these PubMed-induced connections, we constructed 4 additional homogeneous networks through HinMine inspired network decomposition. We used the existence of \emph{inhibition\_TF}, \emph{activation\_TF}, \emph{other\_TF}, \emph{binding} between Arabidopsis Thaliana genes to induce edges between potato genes, each yielding one more homogeneous network.

Finally, as the experimental edges in the heterogeneous network already connect potato genes to other potato genes, we extracted homogeneous networks from them by simply taking only the experimental connections and potato genes from the heterogeneous network, thus yielding the final two homogeneous networks: \emph{experimental\_healthy} and \emph{experimental\_infected}.

\subsection{Feature vector construction}

As most machine learning algorithms are designed to learn from collections of feature vectors (i.e. tabular data), the next step of DDeMON is the construction of such vectors from the 7 homogeneous networks constructed in the previous step. There are many possible ways to obtain real-valued descriptors of individual nodes. 
DDeMON  implements the HinMine approach(see Section~\ref{subs:HINMINE}) where, for each node, a vector of real values of length equal to the total number of nodes is obtained, using the Personalized PageRank (P-PR) network node ranking algorithm. Albeit spatially complex, P-PR ranking demonstrated to produce useful results. 


The P-PR vector of network node $v$ ($\mbox{P-PR}_v$) is defined as the vector representing the probabilities of a stationary distribution of the position of a random walker traversing the nodes in the network. The random walker starts the walk in node $v$. At each step, while residing on node $w$, the random walker then either
\begin{itemize}
    \item with probability $p$, jumps to one of the neighbors of $w$. In this case, the probability that neighbor $n$ of $w$ is selected is equal to $\frac{\mathrm{weight}(w, n)}{\mathrm{deg}(w)}$, where $\mathrm{weight}(w, n)$ denotes the weight of the edge from $w$ to $n$, and $\mathrm{deg}(w)$ is the sum of all weights on outgoing edges of $w$ (i.e. a normalisation constant ensuring probabilities sum to $1$). If the node $w$ has no outgoing edges, then the random walker jumps back to $v$, 
    \item with probability $(1-p)$, jumps back to $v$ and restarts the random walk.
\end{itemize}
\noindent Probability $p$ 
is a parameter of the P-PR algorithm, usually set to $0.85$. 

While $\mbox{P-PR}_v$ is defined as a vector of probability values, where the $i$-th element of the vector $\mbox{P-PR}_v(i)$ is the probability of the walker being at the $i$-th node in the network, each $\mbox{P-PR}_v(i)$  has other equivalent interpretations:
\begin{enumerate}
    \item the proportion of time that the random walker spends at the $i$-th node of the network if the random walk takes place for a long time (i.e. the limit of the expected proportion of time spent as the number of steps is arbitrarily large),
    \item the probability that, after letting the random walk run for a long time, the random walker will be at the $i$-th node of the network,
    \item the limit of a recursively defined sequence of vectors $\{r^{(k)}\}_{k=1}^\infty$, where $r^{0}$ is a vector of all zeroes, except at the position representing node $v$, and 
     $ r^{(k+1)} = p (A^T r^{(k)}) + (1-p) r^{(0)},$
    where $A$ is the coincidence matrix of the network, normalized so that the elements in each row sum to $1$. If all elements in a given row of the coincidence matrix are zero (i.e. if a vertex has no outgoing connections), the column representing node $1$ is set to $1$ (this simulates the behaviour of the walker when jumping from a node with no outgoing connections back to node $v$).
\end{enumerate}

The most useful for calculating the P-PR vectors is the last interpretation (3), repeating iterative 
$r^{(k)}$ calculation until changes in $r^{(k)}$ are sufficiently small.  It can be shown that the construction of the random walk (including a non-zero probability of jumping back to the starting node and the removal of ``dead end'' nodes with no outgoing connections) ensures that the random walk is stochastic (i.e. non-periodic and irreducible) with a transition matrix with the second largest eigenvalue of $p$ (the probability of restarting the walk), ensuring that the iteration of $r^{(k)}$ calculation
will always converge to a single limiting value, with the error reducing by a factor of $p$ after each step. In our experiments, the iteration indeed resulted in quick convergence after only $50$ steps.
Once calculated, the resulting P-PR vectors are normalized according to the Euclidean norm. For each node $v$, the resulting vector  contains information about the proximity of node $v$ to each of the other network nodes. 

\section{DDeMON experiments in predicting gene functions from generated feature vectors}
\label{sec:ddemon-prediction}


We consider the P-PR vectors of a node as a propositionalized feature vector of the node. Because two nodes with similar P-PR vectors will be in the proximity of similar nodes, a classifier should consider them as similar instances. This section presents how the constructed feature vectors for the nodes $v$ are used in network nodes classification.

\subsection{Dimensionality reduction}
\label{subs:dimensionality}

The feature vectors constructed by the P-PR algorithm can in principle be used by any standard machine learning algorithm to provide predictions of any target feature, associated with the potato genes represented by the network nodes. However, each network node appears in all networks, and each feature vector constructed by the P-PR algorithm contains $|V| = 33{,}937$ features (one feature for each node in the network). This means that by concatenating the feature vectors describing the node would result in feature vectors that contain a total of $7\cdot 33{,}937$ features. This high dimensionality makes the data set difficult to analyze with most machine learning algorithms, and additionally, contains too much noise for the algorithms to easily extract the information encoded in the high-dimensional feature vectors.

Our solution to this problem is dimensionality reduction, a technique commonly employed when the dimensionality of the data set is too high. We reduced the dimensionality of each context to dimension $d$ using the standard PCA (Principal Components Analysis) approach.

\noindent This allowed us to reduce the size of the network to a manageable size and ensure that the produced feature vectors can practically be used by several supervised machine learning algorithms in the final learning step.

\subsection{Learning from resulting feature vectors}
\label{subs:learning}

The steps, described so far in this section, yield a set of feature vectors (derived from both expression and network data) that is suitable for all propositional learners. Note that recent improvements in graph representation learning, which include graph-convolutional and similar neural network layers already support similar functionality, yet are prohibitively expensive when global network properties are taken into account. We circumvent this issue by first computing global network properties, followed by dimensionality reduction. Such approach is suitable for \emph{transductive} learning tasks, i.e. tasks where unknown nodes' presence is known, yet their classes are not. 
We tested the following five different state-of-the-art classification algorithms:
\begin{itemize}
\item Support Vector Machines (SVM) - balanced
\item Support Vector Machines (SVM) - adapted to imbalanced data sets
\item Gradient Boosting Machines (GBM)
\item Deep Neural Networks (DNN) - two distinct architectures 
\end{itemize}

As we observed that majority of the classes are imbalanced, we further investigated whether learners' performance can be improved by balancing the data sets according to a given target class. We implemented a simple oversampling procedure, which simply multiplies the instances, labeled with the minority class by a constant factor.

We used all the classification algorithms not only to classify the data, but also to produce a score for each gene-GMM BIN pair. The score (distance from the separating hyperplane in the case of SVM, probability of a positive example in GBM and probability estimate---sigmoid activation---per target) in the case of neural networks) is then used to produce a ranked list of genes for each GMM bin. The list begins with those genes that the classification algorithm estimates are most likely to belong to that bin. This allows us to (1) better estimate the performance of the algorithms by analyzing how accurate they are for different levels of certainty via ROC curves and Precision-Recall (PR) curves, and (2) to produce better results on previously unseen data, as the result ``gene $x$ is in the top $10$ genes most likely to belong to BIN $y$'' holds more informative value than simply a result of ``gene $x$ belongs to BIN $y$''.

\subsection{Validation approach}
\label{subs:validation}

The performance of DDeMON is demonstrated on the task of ontology-based \emph{function prediction}. We consider the following scenarios. 

To validate the method's performance, only the annotated part of the gene network is used, i.e. the genes with the corresponding mappings in the first 34 functional categories in GoMapMan (only the so-called GMM BINs)  \cite{ramvsak2014gomapman} are considered). In BIN 35, all the genes with unknown function are listed: these genes with unknown function will be used in the gene function prediction experiments.

The subset  of functionally annotated genes for potato in the GoMapMan ontology, which are relevant for this task, consists of more than 20{,}000 annotated genes, and is as such a suitable source for the demonstration of the DDeMON's scalability.

The first set of experiments is aimed at answering the question how well DDeMON is able to predict a given gene's functions. In this setting, using cross-validation, parts of the annotated network are hidden during evaluation. The purpose of this experiment is to assess how different representations (layers) of the considered input multiplex network impact the classification performance, when considered alongside different learners, such as deep neural networks or support vector machines.

Once the DDeMON's performance was established using cross-validation, DDeMON was trained using all known gene-annotation associations, and used to predict annotations for the genes with yet unknown functions (BIN 35). As there are no ground truth annotations for these genes, expert analysis was conducted to assess the potential correctness of the newly predicted annotations.

\subsection{Experimental setting}
\label{subs:setting}

The tests performed to evaluate the DDeMON methodology were split into two parts. Section \ref{subs:setting_step1}  describes how the various machine learning algorithms used in the last step of DDeMON were tested, and how the optimal configuration was discovered, while Section~\ref{subs:prediction} describes how the performance of the chosen optimal DDeMON configuration was used in a series of \emph{in silico} validations of DDeMON predictions. 

\subsubsection{Quantitative analysis of DDeMON approach}
\label{subs:setting_step1}

As described in Section \ref{subs:learning}, the DDeMON methodology produces a matrix of predictions, one score for each gene and each GMM bin. In our preliminary testing, we tested the classification algorithms on their ability to predict how likely a gene is to belong to GoMapMan BIN 20.1: biotic stress. We expected good results for classification into this BIN as the initial in vivo experimental setup was designed to activate genes associated with biotic stress.

\begin{itemize}
\item For SVM classifiers, we performed a grid search over $84$ possible parameter configurations:
\\ \noindent -- We used four different SVM kernels: linear, RBF, and polynomial.
\\ \noindent -- For all possible SVM kernels, we varied the $C$ parameter to the values $[10, 1, 0.001, 0.0001]$.
\\ \noindent -- For the RBF and polynomial kernels, we varied the $\gamma$ parameter to the values $[10, 1, 0.001, 0.0001]$.
\\ \noindent -- For the polynomial kernel, we tested degrees of $2$, $3$ and $5$.

\item The GBM's hyperparameters were set as defaults.

\item The hyperparameter settings of the neural learners were as follows. The simpler version of the neural network was parametrized as dense(300)-dropout(0.2)-dense(20)-dense(out), where dense() represents a dense layer and dropout the application of dropout after a given layer. The dense(out) represents the number of output neurons. The larger neural network was parametrized as follows: dense(512)-dropout(0.3)-dense(386)-dropout(0.3)-dense(128)-dropout(0.3)-dense(64)-dropout(0.3)-dense(out).
\end{itemize}


\subsubsection{\textit{In silico} validation of predictions}
\label{subs:prediction}


We evaluated the results using the receiver operating characteristic curves (ROC), with corresponding area under the ROC curve (AUC) scalar values assessing individual binary classifications. On a first glance, the performance of all three classes of classifiers, described in Section~\ref{subs:learning}, is almost identical, as all three achieve similar AUC scores. However, further analysis showed their different performance profiles in terms of the area under the PR curve (AUPR).

\begin{figure}[b!]
    \centering
    \includegraphics[width=\linewidth]{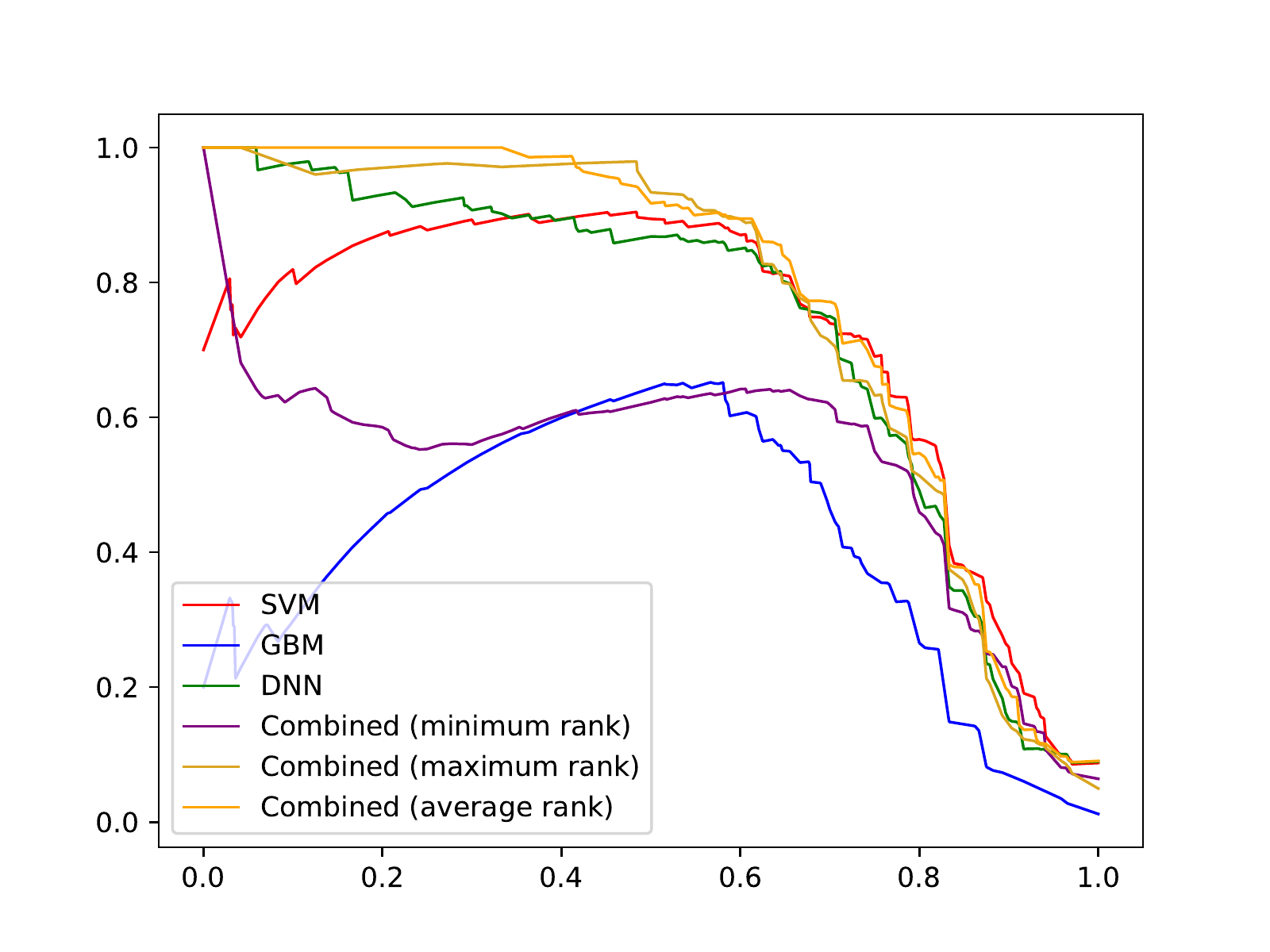}
    \caption{Comparison of best individual, and aggregate classifiers (PR curve), showing that the average rank classifier provides the best recall values. }
    \label{fig:agg}
\end{figure}

Because none of the three classifier types can objectively be said to be better than the other two, we decided to aggregate their results before using them in our \emph{in silico} validation. We constructed aggregate classifiers that use the individual (DNN, SVM and GBM) classifiers to produce---for each GMM BIN---three ranked lists of candidate genes (genes, deemed more likely by a classifier to belong to a BIN, are ranked higher). The aggregate classifiers then produce predictions by taking the average, minimum or maximum ranks of each gene over the three lists. Our experiments (see Figure~\ref{fig:agg}) show that the ``average'' aggregate classifier performed best, with its main advantage being its high precision at high threshold settings (i.e. the left-upper part of the precision-recall curve). The ``average rank'' classifier was then used for all \emph{in silico} validation experiments. 

The DDeMON approach was used to predict involvement of genes with unknown function in a particular pathway, for 10 selected ontology categories (GMM BINs) \cite{ramvsak2014gomapman}. The bins were selected based on the performance of the classifiers. 
The bins in category 35: not assigned were not chosen for further analysis. The 10 top bins, for which the classifiers showed the best performance, and were selected, are:
\begin{itemize}
\item 1.1 PS.lightreaction (281 genes in the training set)
\item 2.1 major CHO metabolism.synthesis (43 genes)
\item 17.5 hormone metabolism.ethylene (217 genes)
\item 17.7 hormone metabolism.jasmonate (58 genes)
\item 17.8 hormone metabolism.salicylic acid (21 genes)
\item 20.1 stress.biotic (1173 genes)
\item 26.8 misc.nitrilases, nitrile lyases, berberine bridge enzymes, reticuline oxidases, troponine reductases (126 genes)
\item 26.22 misc.short chain dehydrogenase/reductase (SDR) (105 genes)
\item 26.28 misc.GDSL-motif lipase (127 genes)
\item 34.13 transport.peptides and oligopeptides (107 genes)
\item 34.16 transport.ABC transporters and multidrug resistance systems (218 genes)
\end{itemize}
From the resulting predictions for each BIN, only genes above a threshold value of 95\% correct for DNN were considered as reliable. InterProScan \cite{quevillon2005interproscan} was run on all the genes of the training and test set to get an independent insight into potential gene function. InterPro and PFAM~\footnote{\url{https://pfam.xfam.org/}} annotations were further used for validation, e.g., to estimate the level of the agreement between annotations of both sets.

For each of the bins listed above, we created a list of candidate genes. The list contained only genes that are currently listed in GMM BIN 35 (unknown). Among these genes, we selected the top $n$ genes as predicted by our classifier, where $n$ was chosen such that on the training data, the top $n$ genes selected by the classifier contained $95\%$ genes that actually belonged to the BIN. 

\section{Experimental results}
\label{sec:results}


\subsection{Results of quantitative analysis}
\label{subs:quantitative}

When testing the $84$ SVM configurations and their applications to classifying genes into BIN 20.1, the results varied significantly, with some SVM configurations performing very well and others performing quite poorly. 
Figures showing the PR curves and the ROC curves of 5 best performing and 5 worst performing SVM configurations are provided in Supplementary material.
The quantitative evaluation of 5 best performing and 5 worst performing SVM configurations is given in Table~\ref{tab:svm}. While the difference between the top performing configurations was small, we decided that for further experiments, the polynomial kernel with settings $C=1, \mathrm{degree}=3, \gamma=10$ shall be used.


\begin{table}
    \caption{Performance of the five best and five worst performing SVM classifiers.}
    \centering
    \begin{tabular}{c|ccc|cc}
     \hline
        Kernel & $C$ & $\gamma$ & Degree  & AUC & AUPR \\\hline
        Polynomial & $1$      & $10$     & $3$ & $0.920$ & $0.475$\\
        Polynomial & $10$     & $1$      & $2$ & $0.919$ & $0.480$\\
        Polynomial & $1$      & $1$      & $2$ & $0.919$ & $0.476$\\
        Polynomial & $01$     & $10$     & $2$ & $0.919$ & $0.476$\\
        Polynomial & $1$      & $10$     & $2$ & $0.919$ & $0.476$\\
        Polynomial & $0.001$  & $0.0001$ & $5$ & $0.501$ & $0.071$\\
        Polynomial & $0.0001$ & $0.0001$ & $5$ & $0.502$ & $0.476$\\
        Polynomial & $1$      & $0.0001$ & $5$ & $0.571$ & $0.421$\\
        RBF        & $1$      & $10$     & /   & $0.749$ & $0.129$\\
        RBF        & $1$      & $10$     & /   & $0.749$ & $0.129$ \\\hline
    \end{tabular}
           \label{tab:svm}
\end{table}

Comparing the SVM results with the results of GBM and DNN, Table \ref{tab:summary} shows the results of the top performing SVM classifier and the DNN and GBM classifiers, while Figures illustrating the SVM results are provided in Supplementary materials.

\begin{table}
    \centering
        \caption{Comparative evaluation of the best SVM, DNN and GBM classifiers.}
    \label{tab:summary}
    \begin{tabular}{ccc}
    \hline
        Classifier& AUC & AUPR \\\hline
        SVM & $0.920$ & $0.475$\\
        GBM & $0.911$ & $0.508$ \\
        DNN & $0.897$ & $0.469$\\
        \hline
    \end{tabular}
\end{table}

\subsection{Results of \emph{in silico} validation}
\label{subs:insilico}

Each gene, predicted to belong to one of the BINs, described in Section~\ref{subs:prediction}, was inspected. Several of the predictions can independently be shown to actually belong to the bins and can therefore be considered successful discoveries of our methodology. An overview of the results is shown in Table~\ref{tab:insilico}. 

The BINs with the largest number of predictions, that are also in concordance with gene annotations of the pathway genes and as expected from the experimental data used, were the following: 20.1 (biotic stress; 19 out of 21 genes) and 34.16 (ABC transporters; 13 out of 18 genes). The remaining BIN predictions were less exact, with a smaller number of predicted genes, however a level of accord between annotations and descriptions was still present: 1.1 (3 out of 4), 17.5 (3 out of 7), 17.7 (2 out of 3), 26.8 (2 out of 2) and 26.28 (3 out of 8). Both predictions for bin 26.22 were discarded, as InterProScan results were not in agreement. The last three BINs (2.1, 17.8 and 34.13) did not result in any gene predictions above our threshold (see Table~\ref{tab:insilico} for details).

\begin{table}[t]
\centering
\caption{The considered bins and the resulting classification performance.}
\resizebox{.5\textwidth}{!}{
\begin{tabular}{ccc}
\hline
BIN ID  & Training  & 95\%  \\ 
 & set size & correct \\\hline
1.1 PS.lightreaction                                                                                         & 281               & 4                                                                                                   \\
2.1 major CHO metabolism.synthesis                                                                           & 43                & 0                                                                                                   \\
17.5 hormone metabolism.ethylene                                                                             & 217               & 7                                                                                                   \\
17.7 hormone metabolism.jasmonate                                                                            & 58                & 3                                                                                                   \\
17.8 hormone metabolism.salicylic acid                                                                       & 21                & 0                                                                                                   \\
20.1 stress.biotic                                                                                           & 1173              & 21                                                                                                  \\
26.8 misc.nitrilases, nitrile lyases, berberine bridge enzymes,  & 126               & 2                                                                                                   \\
reticuline oxidases, troponine reductases &                &                                                                                                 \\
26.22 misc.short chain dehydrogenase/reductase 			(SDR)                                                      & 105               & 4                                                                                                   \\
26.28 misc.GDSL-motif lipase                                                                                 & 127               & 8                                                                                                   \\
34.13 transport.peptides and oligopeptides                                                                   & 107               & 0                                                                                                   \\
34.16 transport.ABC transporters and multidrug 			resistance systems                                         & 218               & 18   \\ \hline                                                                                            
\end{tabular}}
\label{tab:insilico}
\end{table}

\vspace*{-0.5cm}

\section{Discussion and conclusions}
\label{sec:discussion}

One of the key contributions of DDeMON is its capability to directly incorporate the  representation of the entities of interest (e.g., genes in this paper) from the multiple different contexts. In this work, information from both empirical data as well as extensive, freely available background knowledge was incorporated, showing how different sources of knowledge can be jointly considered by a single learning system. Albeit enabling fast representation learning across multiple contexts, the currently main drawback is DDeMON's spatial complexity. As most of the considered experiments were conducted on specialized hardware (GPU cluster with lots of RAM), further developments could go in direction of considering more compact, low-dimensional representations, albeit at the cost of potentially lower performance. Alternatively, only the subset of genes, crucial for the network structure could be used to obtain the relevant representations, also resulting in lower spatial complexity. Finally, even though the experiments in this paper indicate that all modalities need to be considered for maximum performance, we have yet to explore the potential implications of performing feature ranking 
prior to learning.

As part of the proposed method, we explored different learning algorithms, each with their own unique capabilities and language biases. The current results indicate that deeper neural network models (feedforward architecture) outperform shallow models such as SVMs, however, with proper regularization, even SVMs perform competitively, indicating that the representation used as input contains enough information even for simpler models to perform well. The neural network architecture that is employed by DDeMON was designed manually, by incrementally adding hidden layers (and relevant activations), however, the recent advancements in autoML systems, capable of discovering such architectures automatically, could be considered in further work. 

The best performing BIN predictions were, as expected, from BINs 20.1 (biotic stress) and 34.16 (ABC transporters), first a direct result of our chosen dataset and the second important for hormone transport and stress responses~\cite{Kang2011}. The worst performing BINs (2.1, 26.22 and 34.13) were chosen mostly as prediction controls, thus good predictions were not expected given the used biotic stress experimental dataset. Bad prediction performance of BIN 17.8 (salicylic acid) however is contrary to the knowledge of it being a major actor in plant biotic defenses; this however can be explained by the lacking knowledge and annotations of training genes belonging to this BIN.


Overall, the DDeMON approach offers a great simplification of the standard protocol for manual curation of gene ontology descriptions, thus enabling quicker and more exact extraction of knowledge, whilst also incorporating actual experimental data for these predictions.  However, each network node appears in all networks, and each feature vector constructed by the P-PR algorithm contains $|V| = 33{,}937$ features (one feature for each node in the network). 

Currently, we demonstrated the use of DDeMON for a non-model organism, where it offered high-quality predictions, and is a potentially very useful resource. However, if sufficient background knowledge is not available, the obtained node representations are not necessarily of such high quality, making DDeMON perform worse -- such situation could occur when considering a different non-model organism. We believe that automatic incorporation of real-life knowledge graphs potentially built from literature could facilitate such endeavor.

Even though the existing methodology scales to large, real-life biological networks, the following improvements could facilitate its use for even larger data sets. The obtained multi-modal node representations are inherently quadratic with respect to the number of nodes. Such complexity can be prohibitive when considering larger networks, as even for the current version, off-the-shelf hardware was not sufficient (we needed 64GB of RAM). Further work in this direction will consider compressed node representations, making the method subquadratic in space. Finally, the proposed DDeMON could be implemented as a web service-based application, facilitating the use to non-progamming savvy users.



\begin{backmatter}


\section*{Funding}
This work was supported by the Knowledge Technologies national research grant,  and the European Union's Horizon 2020 research and innovation programme under grant agreement 863059 (FNS-Cloud, Food Nutrition Security).




\section*{Competing interests}
The authors declare that they have no competing interests.

\section*{Consent for publication}
All authors agree with publication of this paper.

\section*{Authors' contributions}
B\v{S}, JK and \v{Z}R implemented the approach and conducted the experiments. NL and KG coorodinated the study and suggested improvements to the method/experiment design.



\bibliographystyle{bmc-mathphys} 
\bibliography{bmc_article}      







\end{backmatter}
\end{document}